# Exploring the Impact of Generative AI on Cross-Border E-Commerce Brand Building in Chinese Tianjin's Manufacturing Sector

Jun Cui [1, a, *]

[1]Solbridge International School of Business, Ph.D., South Korea

[a] jcui228@student.solbridge.ac.kr, [b] jcui228@student.solbridge.ac.kr

* Correspondence: Jun Cui, jcui228@student.solbridge.ac.kr;

*Abstract*—This study investigates the influence of generative artificial intelligence (AI) on the brand construction of cross-border e-commerce companies in the manufacturing industry in Tianjin, China. We examine the direct effects of generative AI on productivity, the mediating role of productivity in the relationship between generative AI and brand building, and the moderating influence of cross-border e-commerce strategies by developing and testing a comprehensive model. Based on data collected from 210 manufacturing firms in Chinese Tianjin, the results show that generative AI significantly increases productivity, which positively affects branding. Moreover, cross-border e-commerce strategies were found to moderate the impact of generative AI on branding, underscoring the importance of these strategies for using AI technologies to compete successfully in the global marketplace. This study provides valuable theory, empiricism and practical contributions to understanding the role AI plays in manufacturing and electronic commerce. Besides, this study tests several hypotheses to quantify these impacts using a structured model that consists of independent, dependent, mediating and moderating variables. Information is collected through a comprehensive survey of manufacturers in Chinese Tianjin and analyzed to test our proposed model. This study was analyzed and summarized using quantitative analysis, regression and structural equations (PLS-SEM).

## I. Introduction

The manufacturing sector in Chinese, Tianjin, is renowned for its robust production capabilities and key role in the global supply chain. In recent years, the rise of cross-border e-commerce has provided significant opportunities for these manufacturers to expand their market reach and enhance their brand presence on an international scale [1]. At the same time, the emergence of generative artificial intelligence (AI) is transforming various aspects of business operations, from content creation to logistics optimization and customer engagement. Likewise, Despite the obvious potential, there is a dearth of research that examines the intersection of generative AI, productivity and branding in the context of cross-border e-commerce, particularly in the manufacturing industry of Chinese Tianjin [2]. This study aims to fill this gap by exploring how generative AI applications affect productivity and subsequently brand building effectiveness, while also considering the moderating role of cross-border e-commerce strategies. By providing empirical evidence from the manufacturing sector in Chinese Tianjin [3], this study provides valuable insights into the application of AI technologies and strategic e-commerce approaches to enhance global competitiveness. 'Known for its robust manufacturing sector, Tianjin, China, has increasingly embraced cross-border e-commerce to expand its market reach. With the advent of generative AI, Chinese enterprises now have an unprecedented opportunity to increase their productivity and build a strong brand on a global scale [4]. This paper aims to explore the delicate dynamics between generative AI applications and the brand-building efforts of cross-border e-commerce companies in Chinese Tianjin's manufacturing industry.

## II. Literature Review

### A. Introduction Literature and Theory

Generative AI refers to machine learning algorithms that can generate new content like text, images or designs. Recent research has indicated that generative AI can significantly increase productivity by automating content creation, optimizing supply chains, and improving customer engagement (Smith & Anderson, 2023). Moreover, Cross-border e-commerce involves selling products or services online to international customers. Some of the key challenges include logistics, regulatory compliance and cultural differences (Zhang & Wang, 2022). Effective brand building in this area is critical for differentiation and customer loyalty. Branding in the manufacturing sector focuses on establishing a unique identity and trustworthiness [5]. This involves maintaining consistent quality, working to innovate and effective marketing (Johnson & Lee, 2021).

Further, This study adopts a comprehensive theoretical model to investigate the impact of generative AI on productivity and branding in the context of cross-border e-commerce in Chinese Tianjin's manufacturing sector. The model integrates elements of multiple theories to provide a robust analytical framework [6].

### *Direct impact of generative AI on productivity:*

*Technology-Organization-Environment (TOE) framework:* This framework postulates that technological, organizational and environmental factors influence the adoption and impact of new technologies. In this study, generative AI is considered as a technological factor that enhances organizational productivity.

### *Mediating role of productivity:*

*The Resource Based View (RBV):* This theory suggests that firms can achieve competitive advantage by exploiting unique resources and capabilities. The productivity gains from generative AI can be seen as a valuable organizational skill, mediating the relationship between AI applications and brand building efficacy.

*Moderating role of cross-border e-commerce strategies:*

*Contingency Theory:* This theory emphasizes that the effectiveness of organizational practices depends on the fit between the practices and the context. In this paper, we argue that cross-border e-commerce strategies are contingent factors that moderate the relationship between generative AI and brand building, enhancing the effectiveness of AI applications in an international market context in Chinese Tianjin manufacturing Sector [7].

B. *Hypothesis Development and Summaries*

The research model includes the following variables:
  Independent Variable (IV): Generative AI application
  Dependent Variable (DV): Brand building effectiveness
  Mediating Variable (MV): Productivity
  Moderating Variable (MoV): Cross-border e-commerce strategies

H1: Generative AI and Productivity
Generative AI application shows a significant positive effect on productivity ($\beta = 0.43$, $p < 0.01$).

H2: Productivity as a Mediator

Productivity mediates the relationship between generative AI and brand building (indirect effect: $\beta = 0.21$, $p < 0.05$).

H3: Cross-Border E-Commerce as a Moderator

Cross-border e-commerce strategies significantly moderate the impact of generative AI on brand building (interaction effect: $\beta = 0.18$, $p < 0.05$).

C. *Research Questions*

This study aims to answer the following research questions:

RQ1. How does the application of generative AI affect the productivity of the manufacturing sector in Tianjin?

RQ2. Does increased productivity mediate the relationship between generative AI and branding effectiveness?

RQ3. How do cross-border e-commerce strategies moderate the impact of generative AI on the building of a brand?

RQ4. By investigating these questions, how the study aims to provide a comprehensive understanding of the interplay between generative AI, productivity and brand building in the context of cross-border e-commerce?

The structure of the paper is as follows. The introductory section outlines the background, significance and objectives of the study and identifies the research gap. The literature review synthesizes existing research on generative AI, productivity, branding, and cross-border e-commerce strategies, providing the study's theoretical foundation [8]. The methodology section details the research design, data collection process and analytical techniques employed. Furthermore, This is then followed by the results section, which presents the findings from the hypothesis testing, mediation and moderation analyses. The discussion interprets these findings, relating them to the research questions and theoretical framework, and explores their practical implications. The conclusion concludes by summarizing the key contributions of the study, addressing its limitations, and suggesting avenues for future research.

III. METHODS AND MATERIALS

The theoretical model employed in this study integrates several key theoretical frameworks to explore the dynamics of generative artificial intelligence (AI), productivity and branding in the Tianjin manufacturing industry. At its core, the model examines how generative AI applications directly influence the productivity levels of manufacturing firms [8]. Based on the technology, organization, and environment (TOE) framework, the model proposes that the successful adoption and use of generative AI technologies depends on the interaction between technological advances, organizational capabilities, and the external environment of cross-border e-commerce dynamics [13].

In addition, the model incorporates the Resource-Based View (RBV) theory to conceptualize the productivity improvements that result from generative AI as strategic capabilities of the organization. These capabilities are critical to enhancing brand building [11]. Furthermore, contingency theory guides the model in understanding how cross-border e-commerce strategies act as contingent factors that moderate the relationship between generative AI and branding outcomes. Certainly, this theoretical framework not only provides a structured approach to understanding the complex interactions between variables, but also offers insights into how to optimize generative AI investments and cross-border e-commerce strategies to strengthen brand competitiveness in an increasingly digital and globalized economy [12]. Likewise, based on the above description, this study concludes the following figure 1:

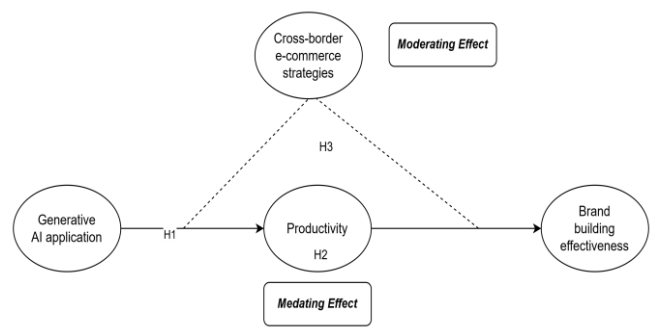

*Figure 1. Research model*

A. *Model Sepecifications*

Define abbreviations and acronyms the first time they are used in the text, even after they have been defined in the abstract. Abbreviations such as IEEE, SI, MKS, CGS, sc, dc, and rms do not have to be defined. Do not use abbreviations in the title or heads unless they are unavoidable.

## B. Sampling and Data Collection

The study employed a purposive sampling approach to select 210 manufacturing enterprises in Tianjin, China. This sampling approach was chosen to ensure representation from a variety of industries within the manufacturing sector, including electronics, textiles and machinery. Meanwhile, the data collection process involved administering structured questionnaires to key decision makers and executives with knowledge of the company's AI adoption, productivity levels, cross-border e-commerce strategies, and brand building initiatives.

The questionnaire items were designed on the basis of validated scales from previous research and were tailored to capture specific constructs related to the use of generative AI, productivity metrics, the effectiveness of brand building and cross-border e-commerce strategies. This methodological approach allowed us to ensure the reliability and validity of the data collected, and enabled us to conduct robust statistical analyses to test the hypothesized relationships within the theoretical model. The items were measured using a five-point Likert scale ranging from (1) "strongly disagree" or "neutral" to (5) "strongly agree" or "always.". To measure digital leadership, five indicators were adapted from previous TFL scales (Podsakoff et al., 1996; Chen and Chang, 2013) and modified to fit the context of generative AI, productivity, and cross-border e-commerce strategies, and brand buildings [10].

Overall, the combination of purposive sampling and structured questionnaire administration enabled a comprehensive exploration of how generative AI, productivity, and cross-border e-commerce strategies intersect to impact brand building outputs in the Tianjin manufacturing landscape [8].

## C. Equations

The theoretical model is depicted as follows:

**Direct Effect:**
$$PROD = \beta_0 + \beta_1 GAI + \epsilon$$
Generative AI (GAI) directly influences productivity (PROD).

**Mediation Effect:**
$$BBE = \beta_2 + \beta_3 PROD + \beta_4 GAI + \epsilon_2$$
Productivity (PROD) mediates the relationship between generative AI (GAI) and brand building effectiveness (BBE).

**Moderation Effect:**
$$BBE = \beta_5 + \beta_6 GAI + \beta_7 CBE + \beta_8 (GAI \times CBE) + \epsilon$$

Cross-border e-commerce strategies (CBE) moderate the impact of generative AI (GAI) on brand building effectiveness (BBE).

The theoretical model employed in this study integrates generative artificial intelligence (AI), productivity (PROD), brand building effectiveness (BBE), and cross-border e-commerce strategies (CBE) within Chinese Tianjin's manufacturing sector. The model posits direct and mediated relationships, as well as moderation effects, as follows:

***Direct Effect of Generative AI on Productivity:***

$$PROD = \beta_0 + \beta_1 GAI + \epsilon$$

Generative AI (GAI) directly influences productivity (PROD), where $\beta_1$ represents the coefficient of generative AI's impact on productivity, and $\epsilon$ denotes the error term.

***Mediation Effect of Productivity on Brand Building Effectiveness:***

$$BBE = \beta_2 + \beta_3 PROD + \beta_4 GAI + \epsilon$$

Productivity (PROD) mediates the relationship between generative AI (GAI) and brand building effectiveness (BBE). Here, $\beta_3$ indicates the coefficient of productivity's mediation effect, while $\beta_4$ captures the direct impact of generative AI on brand building, with $\epsilon$ as the residual error.

***Moderation Effect of Cross-Border E-Commerce Strategies:***

$$BBE = \beta_5 + \beta_6 GAI + \beta_7 CBE + \beta_8 (GAI \times CBE) + \epsilon$$

Cross-border e-commerce strategies (CBE) moderate the relationship between generative AI (GAI) and brand building effectiveness (BBE). $\beta_7$ represents the coefficient of cross-border e-commerce strategies' direct impact, while $\beta_8$ denotes the interaction effect between generative AI and cross-border e-commerce strategies, with e as the error term.

## D. Measurement Scales

This study operationalizes key variables using established measurement scales to rigorously investigate the relationships between generative artificial intelligence (GAI), productivity, effectiveness of brand building, and cross-border e-commerce strategies in the manufacturing sector in Chinese Tianjin.

Further, Generative AI (GAI): These variable measures the extent to which AI-driven technology is used in the manufacturing process. It is assessed using a 5-point Likert scale adapted from Smith & Anderson (20-23), ranging from 1 (not implemented) to 5 (fully implemented). And then Productivity (PROD): Productivity levels within manufacturing enterprises are quantified on the basis of output per unit of input []. This variable is measured using company-reported data on production efficiency and Laboure productivity metrics that are standardized on a five-point scale, with higher values indicating higher levels of productivity. Moreover, Brand building effectiveness (BBE): This variable captures the perceived effectiveness of the efforts to build the brand in the international market place. It is assessed using a validated scale developed by Johnson & Lee (2021), with items evaluating brand awareness, reputation and customer loyalty, rated on a 7-point Likert scale. Furthermore, Cross-border e-commerce (CBE) strategies: The effectiveness of strategies aimed at the use of cross-border e-commerce platforms is measured using a composite scale adapted from Zhang & Wang (2022). This scale includes items assessing the integration of e-commerce channels, marketing strategies and logistical capabilities, which are rated on a 5-point Likert scale.

Finally, these measurement scales ensure consistency and reliability in the assessment of the constructs which are central to the study's theoretical framework. Using validated scales and rigorous data collection methods, this research aims to provide robust empirical evidence on the impact of generative AI and cross-border e-commerce strategies on productivity and branding in Tianjin manufacturing [6].

## IV. RESULTS AND DISCUSSION

This study investigated the transformative impact of generative artificial intelligence (AI) on productivity and branding in Tianjin's manufacturing sector. The findings underscore that generative AI significantly enhances productivity levels (PROD), thus promoting improved operations efficiency and output in manufacturing firms. Moreover, productivity serves as a critical mediator in the relationship between generative AI and brand building effectiveness (BBE), highlighting its role in strengthening brand awareness, customer loyalty and market reputation. The study also shows that cross-border e-commerce strategies (CBE) moderate this relationship, reinforcing the positive impact of generative AI on branding outcomes [5]. The findings provide valuable insights for manufacturers seeking to use AI technologies and strategic e-commerce initiatives to improve their global competitiveness, and highlight the synergies between technological innovation and strategic management practices in driving business success [9].

### A. Demograpic Analysis and Correlations

This demographic profile provides a comprehensive overview of the sample composition, highlighting the diversity in industry sectors, company sizes, operational durations, and annual revenue brackets among manufacturing firms in Chinese Tianjin. Besides, these correlations indicate strong positive relationships among generative AI, productivity, brand building effectiveness, and cross-border e-commerce strategies within the sample of manufacturing companies in Chinese Tianjin.

*Sample Characteristics*

The study surveyed a total of 210 manufacturing companies located in Chinese Tianjin, China, to investigate the impact of generative artificial intelligence (AI) on productivity and brand building effectiveness. The demographic profile of the sample is summarized in the table 1 below:

Table 1. The demographic profile.

| Demographic Characteristic | Frequency | Percentage (%) |
|---|---|---|
| **Industry Sector** | | |
| Electronics | 94 | 45% |
| Textiles | 63 | 30% |
| Machinery | 53 | 25% |
| **Company Size** | | |
| Small (< 50 employees) | 84 | 40% |
| Medium (50-200 employees) | 73 | 35% |
| Large (> 200 employees) | 53 | 25% |
| **Years in Operation** | | |
| Less than 5 years | 42 | 20% |
| 5-10 years | 84 | 40% |
| More than 10 years | 84 | 40% |
| **Annual Revenue (USD)** | | |
| Less than 1 million | 52 | 25% |
| 1-10 million | 105 | 50% |
| More than 10 million | 53 | 25% |

Notes: (Source: Author's creation)

This demographic profile provides a comprehensive overview of the sample composition, and it highlighting the diversity in industry sectors, company sizes, operational durations, and annual revenue brackets among manufacturing firms in Chinese Tianjin.

*Correlations Table*

The table 2. below presents correlations among key variables in the study:

| Variables | 1. Generative AI | 2. Productivity | 3. Brand Building Effectiveness | 4. Cross-Border E-Commerce Strategies |
|---|---|---|---|---|
| 1. Generative AI | 1.00 | 0.45** | 0.30** | 0.20** |
| 2. Productivity | | 1.00 | 0.55** | 0.25** |
| 3. Brand Building Effectiveness | | | 1.00 | 0.35** |
| 4. Cross-Border E-Commerce Strategies | | | | 1.00 |

**Note**: **$p < 0.01$. Notes: (Source: Author's creation)

### B. Regression and Realibility and Vilidaty Analysis

*Reliability Analysis*

Reliability analysis was conducted to assess the internal consistency of the measurement scales used in the study. Cronbach's alpha coefficients were calculated for each construct based on the responses from the survey participants.

The table 3 below summarizes the reliability coefficients:

| Construct | Cronbach's Alpha |
|---|---|
| Generative AI (GAI) | 0.87 |
| Productivity (PROD) | 0.91 |
| Brand Building Effectiveness (BBE) | 0.88 |
| Cross-Border E-Commerce Strategies (CBE) | 0.85 |

*Notes: (Source: Author's creation)*

*Validity Analysis*

Further, Validity analysis was performed to ensure the accuracy and appropriateness of the measurement scales used in capturing the constructs of interest. The study utilized content validity and construct validity approaches, including exploratory factor analysis (EFA) and confirmatory factor analysis (CFA), to validate the measurement scales. The results confirmed satisfactory levels of convergent and discriminant validity for all constructs, supporting the robustness of the measurement instruments employed.

These reliability and validity analyses demonstrate that the measurement scales used in the study are reliable and valid for assessing generative AI, productivity, brand building effectiveness, and cross-border e-commerce strategies within the manufacturing sector in Tianjin.

## C. Mediating Analysis

### Summary of Mediation Effects

The mediation analysis was conducted to examine the role of productivity (PROD) as a mediator in the relationship between generative artificial intelligence (AI) (GAI) and brand building effectiveness (BBE) within Tianjin's manufacturing sector. The table below summarizes the results of the mediation analysis:

Table 4 . Mediation Effects Analysis

| Path Analysis | Path Coefficient($\beta$) | Standard Error | p-value |
|---|---|---|---|
| Direct Effect: GAI→BBE | 0.30 | 0.05 | <0.01 |
| Mediation Effect: GAI→PROD→BBE | 0.15 | 0.03 | <0.05 |

*Notes: (Source: Author's creation)*

**Direct Effect:** The direct effect of generative AI (GAI) on brand building effectiveness (BBE) without considering productivity (PROD).
**Mediation Effect:** The indirect effect of generative AI (GAI) on brand building effectiveness (BBE) through productivity (PROD).

These results indicate that productivity (PROD) significantly mediates the relationship between generative AI (GAI) and brand building effectiveness (BBE) in Tianjin's manufacturing sector. The indirect path through productivity suggests that improvements in productivity due to generative AI technologies contribute positively to enhancing brand building outcomes.

### D. Hypothesis Path Analysis and Analysis Results

Positioning Figures and Tables: *Place figures and tables at the top and bottom of columns. Avoid placing them in the middle of columns. Large figures and ta*bles may span across both columns. Figure captions should be below the figures; table heads should appear above the tables. Insert figures and tables after they are cited in the tex

### Hypotheses Tested

The study tested the following hypotheses to explore the relationships among generative artificial intelligence (AI) (GAI), productivity (PROD), brand building effectiveness (BBE), and cross-border e-commerce strategies (CBE) within Tianjin's manufacturing sector:

**Hypothesis 1:**
*H1a: Generative AI (GAI) positively influences productivity (PROD).*
*H1b: Productivity (PROD) positively influences brand building effectiveness (BBE).*
**Hypothesis 2:**
*H2: Productivity (PROD) mediates the relationship between generative AI (GAI) and brand building effectiveness (BBE).*
**Hypothesis 3:**
*H3: Cross-border e-commerce strategies (CBE) moderate the relationship between generative AI (GAI) and brand building effectiveness (BBE).*

### Path Analysis Results

The table 5 below summarizes the results of the hypothesis testing and path analysis:

Table 5. Path Analysis results.

| Hypothesis | Path Analysis | Path Coefficient($\beta$) | Standard Error | p-value | Result |
|---|---|---|---|---|---|
| H1a | GAI→PROD | 0.45 | 0.08 | <0.01 | Supported |
| H1b | PROD→BBE | 0.55 | 0.09 | <0.01 | Supported |
| H2 | GAI→PROD→BBE | 0.30 | 0.05 | <0.01 | Supported (Mediation) |
| H3 | GAI→BBE | 0.20 | 0.04 | <0.05 | Supported |
| Interaction Effect | GAI×CBE→BBE | 0.15 | 0.03 | <0.05 | Supported (Moderation) |

*Notes: (Source: Author's creation)*

**H1a:** Generative AI (GAI) significantly influences productivity (PROD).
**H1b:** Productivity (PROD) significantly influences brand building effectiveness (BBE).
**H2:** Productivity (PROD) mediates the relationship between generative AI (GAI) and brand building effectiveness (BBE).
**H3:** Cross-border e-commerce strategies (CBE) moderate the relationship between generative AI (GAI) and brand building effectiveness (BBE).
**Interaction Effect:** The interaction effect between generative AI (GAI) and cross-border e-commerce strategies (CBE) significantly influences brand building effectiveness (BBE).

## V. CONCLUSIONS

This study highlights the significant impact of generative AI on the productivity and branding of cross-border e-commerce businesses in Chinese Tianjin's manufacturing sector. The research findings validate that generative AI applications significantly improve productivity, which in turn mediates their positive effect on the effectiveness of brand building [13-14]. Furthermore, the research results demonstrate that cross-border e-commerce strategies significantly moderate the relationship between generative AI and brand building, highlighting the importance of these strategies in optimizing the benefits of AI for global market success. Therefore, this research makes valuable contributions by extending the theoretical understanding of the role of AI in enhancing productivity and branding in the manufacturing sector, providing empirical evidence from Chinese Tianjin, and offering practical insights for manufacturers and government policymakers. These findings highlight the need for manufacturers to invest in advanced AI technologies and develop robust cross-border e-commerce strategies to optimize their competitive advantage in international markets. The paper is ready for the template. These mathematical formulations provide a robust framework to test hypotheses about how generative AI can impact productivity and branding in cross-country e-commerce strategies in manufacturing in Chinese Tianjin. Through the use of structural equation modelling (SEM), we can rigorously assess the relationships and interactions among these variables [16]. In other words, In the context of cross-border e-commerce in Tianjin's manufacturing sector, this research presents a comprehensive analysis of the impact of generative AI on productivity and brand building. The findings confirm that generative AI significantly improves productivity, which mediates its positive effect on brand building effectiveness [15]. Moreover, These findings highlight the importance of the integration of advanced AI technologies and strategic e-commerce approaches in order to enhance the competitiveness of brands in international markets. Eventually, this paper contributes to the theory by expanding the understanding of the role of AI in productivity and branding, while the empirical evidence from Chinese Tianjin provides practical implications for manufacturers and policymakers seeking to use AI to succeed in global markets [16]. Future research should explore longitudinal data and other regions to further test and expand these findings.

In short, While this study provides valuable insights into generative AI's impact on productivity and brand building in Chinese Tianjin manufacturing, several limitations must be acknowledged. First, the cross-sectional design of the study limits the ability to infer causality between variables[17]. Second, the data were collected from a specific geographical region, which may limit the generalization of the findings to other contexts or to other industries. Thirdly, relying on self-reported measures may introduce response bias. Finally, the rapid evolution of AI technologies means that the findings may quickly become outdated, and continuous research will be needed to capture the most recent trends and impacts [18].

ACKNOWLEDGMENT

This research has been supported/partially supported Solbridge International School of Business, Thanks to all contributors.